\RequirePackage{lineno}
\documentclass[prl,twocolumn,superscriptaddress,showpacs,amsmath,amsfonts,amssymb]{revtex4}
\usepackage{comment}
\excludecomment{printrobustnotes}
\excludecomment{printallnotes}

\usepackage{ifpdf}
\ifpdf
  \usepackage[pdftex]{graphicx}
  \usepackage{epstopdf}
\else
  \usepackage[dvips]{graphicx}
\fi

\setkeys{Gin}{width=.99\columnwidth, height=.75\textheight, keepaspectratio}
\usepackage{dcolumn}% Align table columns on decimal point
\usepackage{bm}% bold math
\usepackage{url}
\usepackage{xspace}

\newenvironment{cfigure}[1][tbp]{\begin{figure}[#1]\centering}{\end{figure}}

\newcommand{\fig}[1]{Fig.~\ref{#1}}
\newcommand{\tab}[1]{Table~\ref{#1}}

\newcommand{\Fig}[1]{Figure~\ref{#1}}
\newcommand{\Tab}[1]{Table~\ref{#1}}

\newcommand{\note}[1]{\xspace}
\newcommand{\robustnote}[1]{\xspace}
\begin{printrobustnotes}
\renewcommand{\robustnote}[1]{\textbf{\textit{#1}}\xspace}
\end{printrobustnotes}
\begin{printallnotes}
\renewcommand{\note}[1]{\textbf{\textit{#1}}\xspace}
\renewcommand{\robustnote}[1]{\textbf{\textit{#1}}\xspace}
\end{printallnotes}
\makeatletter
\@addtoreset{equation}{section}
\renewcommand{\theequation}{\thesection.\@arabic\c@equation}
\makeatother

\newcommand{\alphas}    {\ensuremath{\alpha_{s}}\xspace}

\newcommand{\twotag}  {2-tag\xspace}
\newcommand{\onetag}  {1-tag\xspace}

\newcommand{\ppbar}  {\ensuremath{p\bar{p}}\xspace}
\newcommand{\ttbar}  {\ensuremath{t\bar{t}}\xspace}

\newcommand{\et}     {\ensuremath{E_{T}}\xspace}

\newcommand{\pt}     {\ensuremath{p_{T}}\xspace}

\newcommand{\met}    {\mbox{$\protect \raisebox{.3ex}{$\not$}\et$}\xspace}

\newcommand{\wjets}   {\ensuremath{W~+~\text{jets}}\xspace}
\newcommand{\chisq}   {\ensuremath{\chi^{2}}\xspace}

\newcommand{\mtop}    {\ensuremath{m_{t}}\xspace}
\newcommand{\gtop}    {\ensuremath{\Gamma_{t}}\xspace}
\newcommand{\gtopfit} {\ensuremath{\Gamma_{t}^{\text{fit}}}\xspace}
\newcommand{\mreco}   {\ensuremath{m_{t}^{\text{reco}}}\xspace}

\newcommand{\genunit}[2]{\ensuremath{#1~\mathrm{#2}}\xspace}

\newcommand{\tev}[1] {\ensuremath{#1~\mathrm{TeV}}}
\newcommand{\gev}[1] {\ensuremath{#1~\mathrm{GeV}}}
\newcommand{\gevnoarg}{\ensuremath{\mathrm{GeV}}}

\newcommand{\gevcnoarg}{\ensuremath{\mathrm{GeV}/c}}
\newcommand{\gevcc}[1]{\ensuremath{#1~\mathrm{GeV}/c^{2}}}

\newcommand{\invpb}[1]{\ensuremath{#1~\mathrm{pb}^{-1}}}

\newcommand{\measErr}[2]{\ensuremath{#1 \pm #2}\xspace}

\begin{document}

\title{First Direct Bound on the Total Width of the Top Quark in \ppbar Collisions at $\sqrt{s} = 1.96$~TeV}
\affiliation{Institute of Physics, Academia Sinica, Taipei, Taiwan 11529, Republic of China} 
\affiliation{Argonne National Laboratory, Argonne, Illinois 60439} 
\affiliation{University of Athens, 157 71 Athens, Greece} 
\affiliation{Institut de Fisica d'Altes Energies, Universitat Autonoma de Barcelona, E-08193, Bellaterra (Barcelona), Spain} 
\affiliation{Baylor University, Waco, Texas  76798} 
\affiliation{Istituto Nazionale di Fisica Nucleare Bologna, $^t$University of Bologna, I-40127 Bologna, Italy} 
\affiliation{Brandeis University, Waltham, Massachusetts 02254} 
\affiliation{University of California, Davis, Davis, California  95616} 
\affiliation{University of California, Los Angeles, Los Angeles, California  90024} 
\affiliation{University of California, San Diego, La Jolla, California  92093} 
\affiliation{University of California, Santa Barbara, Santa Barbara, California 93106} 
\affiliation{Instituto de Fisica de Cantabria, CSIC-University of Cantabria, 39005 Santander, Spain} 
\affiliation{Carnegie Mellon University, Pittsburgh, PA  15213} 
\affiliation{Enrico Fermi Institute, University of Chicago, Chicago, Illinois 60637} 
\affiliation{Comenius University, 842 48 Bratislava, Slovakia; Institute of Experimental Physics, 040 01 Kosice, Slovakia} 
\affiliation{Joint Institute for Nuclear Research, RU-141980 Dubna, Russia} 
\affiliation{Duke University, Durham, North Carolina  27708} 
\affiliation{Fermi National Accelerator Laboratory, Batavia, Illinois 60510} 
\affiliation{University of Florida, Gainesville, Florida  32611} 
\affiliation{Laboratori Nazionali di Frascati, Istituto Nazionale di Fisica Nucleare, I-00044 Frascati, Italy} 
\affiliation{University of Geneva, CH-1211 Geneva 4, Switzerland} 
\affiliation{Glasgow University, Glasgow G12 8QQ, United Kingdom} 
\affiliation{Harvard University, Cambridge, Massachusetts 02138} 
\affiliation{Division of High Energy Physics, Department of Physics, University of Helsinki and Helsinki Institute of Physics, FIN-00014, Helsinki, Finland} 
\affiliation{University of Illinois, Urbana, Illinois 61801} 
\affiliation{The Johns Hopkins University, Baltimore, Maryland 21218} 
\affiliation{Institut f\"{u}r Experimentelle Kernphysik, Universit\"{a}t Karlsruhe, 76128 Karlsruhe, Germany} 
\affiliation{Center for High Energy Physics: Kyungpook National University, Daegu 702-701, Korea; Seoul National University, Seoul 151-742, Korea; Sungkyunkwan University, Suwon 440-746, Korea; Korea Institute of Science and Technology Information, Daejeon, 305-806, Korea; Chonnam National University, Gwangju, 500-757, Korea} 
\affiliation{Ernest Orlando Lawrence Berkeley National Laboratory, Berkeley, California 94720} 
\affiliation{University of Liverpool, Liverpool L69 7ZE, United Kingdom} 
\affiliation{University College London, London WC1E 6BT, United Kingdom} 
\affiliation{Centro de Investigaciones Energeticas Medioambientales y Tecnologicas, E-28040 Madrid, Spain} 
\affiliation{Massachusetts Institute of Technology, Cambridge, Massachusetts  02139} 
\affiliation{Institute of Particle Physics: McGill University, Montr\'{e}al, Canada H3A~2T8; and University of Toronto, Toronto, Canada M5S~1A7} 
\affiliation{University of Michigan, Ann Arbor, Michigan 48109} 
\affiliation{Michigan State University, East Lansing, Michigan  48824}
\affiliation{Institution for Theoretical and Experimental Physics, ITEP, Moscow 117259, Russia} 
\affiliation{University of New Mexico, Albuquerque, New Mexico 87131} 
\affiliation{Northwestern University, Evanston, Illinois  60208} 
\affiliation{The Ohio State University, Columbus, Ohio  43210} 
\affiliation{Okayama University, Okayama 700-8530, Japan} 
\affiliation{Osaka City University, Osaka 588, Japan} 
\affiliation{University of Oxford, Oxford OX1 3RH, United Kingdom} 
\affiliation{Istituto Nazionale di Fisica Nucleare, Sezione di Padova-Trento, $^u$University of Padova, I-35131 Padova, Italy} 
\affiliation{LPNHE, Universite Pierre et Marie Curie/IN2P3-CNRS, UMR7585, Paris, F-75252 France} 
\affiliation{University of Pennsylvania, Philadelphia, Pennsylvania 19104} 
\affiliation{Istituto Nazionale di Fisica Nucleare Pisa, $^q$University of Pisa, $^r$University of Siena and $^s$Scuola Normale Superiore, I-56127 Pisa, Italy} 
\affiliation{University of Pittsburgh, Pittsburgh, Pennsylvania 15260} 
\affiliation{Purdue University, West Lafayette, Indiana 47907} 
\affiliation{University of Rochester, Rochester, New York 14627} 
\affiliation{The Rockefeller University, New York, New York 10021} 

\affiliation{Istituto Nazionale di Fisica Nucleare, Sezione di Roma 1, $^v$Sapienza Universit\`{a} di Roma, I-00185 Roma, Italy} 

\affiliation{Rutgers University, Piscataway, New Jersey 08855} 
\affiliation{Texas A\&M University, College Station, Texas 77843} 
\affiliation{Istituto Nazionale di Fisica Nucleare Trieste/\ Udine, $^w$University of Trieste/\ Udine, Italy} 
\affiliation{University of Tsukuba, Tsukuba, Ibaraki 305, Japan} 
\affiliation{Tufts University, Medford, Massachusetts 02155} 
\affiliation{Waseda University, Tokyo 169, Japan} 
\affiliation{Wayne State University, Detroit, Michigan  48201} 
\affiliation{University of Wisconsin, Madison, Wisconsin 53706} 
\affiliation{Yale University, New Haven, Connecticut 06520} 
\author{T.~Aaltonen}
\affiliation{Division of High Energy Physics, Department of Physics, University of Helsinki and Helsinki Institute of Physics, FIN-00014, Helsinki, Finland}
\author{J.~Adelman}
\affiliation{Enrico Fermi Institute, University of Chicago, Chicago, Illinois 60637}
\author{T.~Akimoto}
\affiliation{University of Tsukuba, Tsukuba, Ibaraki 305, Japan}
\author{M.G.~Albrow}
\affiliation{Fermi National Accelerator Laboratory, Batavia, Illinois 60510}
\author{B.~\'{A}lvarez~Gonz\'{a}lez}
\affiliation{Instituto de Fisica de Cantabria, CSIC-University of Cantabria, 39005 Santander, Spain}
\author{S.~Amerio$^u$}
\affiliation{Istituto Nazionale di Fisica Nucleare, Sezione di Padova-Trento, $^u$University of Padova, I-35131 Padova, Italy} 

\author{D.~Amidei}
\affiliation{University of Michigan, Ann Arbor, Michigan 48109}
\author{A.~Anastassov}
\affiliation{Northwestern University, Evanston, Illinois  60208}
\author{A.~Annovi}
\affiliation{Laboratori Nazionali di Frascati, Istituto Nazionale di Fisica Nucleare, I-00044 Frascati, Italy}
\author{J.~Antos}
\affiliation{Comenius University, 842 48 Bratislava, Slovakia; Institute of Experimental Physics, 040 01 Kosice, Slovakia}
% ...

\collaboration{CDF Collaboration\footnote{With visitors from $^a$Universiteit Antwerpen, B-2610 Antwerp, Belgium, 
$^b$Chinese Academy of Sciences, Beijing 100864, China, 
$^c$University of Bristol, Bristol BS8 1TL, United Kingdom, 
$^d$University of California Irvine, Irvine, CA  92697, 
$^e$University of California Santa Cruz, Santa Cruz, CA  95064, 
$^f$Cornell University, Ithaca, NY  14853, 
$^g$University of Cyprus, Nicosia CY-1678, Cyprus, 
$^h$University College Dublin, Dublin 4, Ireland, 
$^i$University of Edinburgh, Edinburgh EH9 3JZ, United Kingdom, 
$^j$Universidad Iberoamericana, Mexico D.F., Mexico, 
$^k$University of Manchester, Manchester M13 9PL, England, 
$^l$Nagasaki Institute of Applied Science, Nagasaki, Japan, 
$^m$University de Oviedo, E-33007 Oviedo, Spain, 
$^n$Queen Mary, University of London, London, E1 4NS, England, 
$^o$Texas Tech University, Lubbock, TX  79409, 
$^p$IFIC(CSIC-Universitat de Valencia), 46071 Valencia, Spain,
$^x$Royal Society of Edinburgh/Scottish Executive Support Research Fellow,
$^y$Istituto Nazionale di Fisica Nucleare, Sezione di Cagliari, 09042 Monserrato (Cagliari), Italy 
}}
\noaffiliation

%\collaboration{CDF Collaboration}
%\noaffiliation
%\date{\today}

\begin{abstract}
We present the first direct experimental bound on the total decay width of the top quark, \gtop, using \invpb{955} of the Tevatron's \ppbar collisions recorded by the Collider Detector at Fermilab.  We identify 253 top-antitop pair candidate events.
The distribution of reconstructed top quark mass from these events is fitted to templates representing different values of the top quark width.  Using a confidence interval based on likelihood ratio ordering, we extract an upper limit at 95\% C.L. of $\gtop<\gev{13.1}$ for an assumed top quark mass of \gevcc{175}.
\end{abstract}

% activate the following line for publication

\maketitle

Due to its large mass, the top quark has the largest decay width and hence the shortest lifetime of the quarks in the standard model (SM). The total width of the top quark at the leading order is dependent on the top quark mass \mtop and the Fermi coupling constant \text{$G_{F}$}: $\gtop^{0}~=~G_{F}\mtop^{3}/(8\pi\sqrt{2})$. Higher order effects include the introduction of finite $W$ boson and $b$ quark masses ($M_{W}, m_{b}$), non-zero off-diagonal elements of the quark-mixing matrix, and higher order corrections in the strong coupling constant \alphas. Neglecting terms of order $m_{b}^{2}/m_{t}^{2}$, $\alphas^{2}$, and $(\alphas/\pi)M_{W}^{2}/m_{t}^{2}$, the width predicted in the SM at next-to-leading-order is:
\begin{equation*}
\gtop = \gtop^{0} (1-\frac{M_{W}^{2}}{m_{t}^{2}})^{2}(1+2\frac{M_{W}^{2}}{m_{t}^{2}})[1-\frac{2\alphas}{3\pi}(\frac{2\pi^{2}}{3}-\frac{5}{2})].
\end{equation*}
 The total width of the top quark is calculated to a precision of about $1\%$ in the SM; it is approximately \gev{1.5} for $m_{t} = \gevcc{175}$~\cite{Czarnecki:1998qc,Chetyrkin:1999ju}.

A deviation from the SM prediction could indicate a significant contribution from top quark decays to non--SM particles such as $t\rightarrow bH^{+}$ (where $H^{+}$ is the charged Higgs boson in the supersymmetric model), or from rare SM processes such as $t\rightarrow dW^{+}$ and $t\rightarrow sW^{+}$. Although such scenarios have not been observed experimentally~\cite{Carena:2000, Abe:1998, Aaltonen:2008, Acosta:2005, Abazov:2008}, a general way to rule out the presence of a large top quark decay rate to non--SM channels, including those with non-detectable final states, is through experimental constraints on \gtop.  To date, there have been no direct experimental measurements of the total width of the top quark.

The data set for the analysis presented in this paper is collected by the CDF II detector, a multipurpose particle detector for \ppbar collisions at $\sqrt{s}=\tev{1.96}$ at the Fermilab Tevatron.  A charged particle tracking system immersed in a magnetic field consists of a silicon microstrip tracker and a drift chamber. Electromagnetic and hadronic calorimeters surround the tracking system and measure particle energies. Drift chambers and scintillators located outside the calorimeters detect muons.  The detector is described in detail elsewhere~\cite{r_cdfDetector}.

We employ a cylindrical coordinate
system where $\theta$\ and $\phi$\ are the polar and azimuthal angle, 
respectively, with respect to the proton beam. Transverse energy and momentum are 
$E_T= E\sin\theta$ and $p_T=|p|\sin\theta$, respectively, where $E$ and
$p$ are energy and momentum.
Missing transverse energy, \met, is defined as the magnitude of the vector
$-\Sigma_{i}\et^{i}\mathbf{n_{i}}$ where $\et^{i}$ is the magnitude of transverse energy contained in each calorimeter tower $i$, and $\mathbf{n_{i}}$ is the unit vector from the collision point to the tower in the transverse plane.

Top quarks are produced primarily by strong interaction in top-antitop (\ttbar) pairs at the Tevatron.  Top quarks decay almost exclusively to a $W$ boson and a $b$ quark through the weak interaction in the SM.
We identify candidate \ttbar events in the ``lepton + jets'' channel, where one $W$ boson decays to an electron or a muon, and a neutrino, while the other $W$ boson decays to a quark-antiquark pair.  We select events consistent with this topology, requiring a high-\pt electron or muon candidate, missing transverse energy denoting the presence of a neutrino, and four jets. Jets are reconstructed using a cone algorithm with radius $\Delta \text{R} \equiv \sqrt[]{\Delta \eta^{2} + \Delta \phi^{2}} = 0.4$. At least one jet with $\et>\gev{15}$ must be identified as a $b$ quark candidate through the presence of a displaced vertex within the jet cone arising from the decay of a long-lived bottom hadron ($b$-tag). The event selection criteria are listed in \Tab{t:evsel}. Detailed information on event selection is available elsewhere~\cite{WZXSec,Acosta:2004hw}.

We divide the candidate events into two exclusive classes: one (\onetag) containing events with one $b$-tagged jet among the leading four, and another (\twotag) with two or more such jets.  Separating these subsamples results in a more efficient use of statistical information due to their different reconstructed top mass resolution and signal-to-background ratios. Background events are expected primarily from $W$ production in association with jets ($W$~+~jets), multijet processes where a jet is misidentified as a charged lepton and \met results from energy mismeasurement of the jets (non-$W$), and small contributions from electroweak backgrounds (EWK) composed of single top quark and diboson ($WW$, $WZ$, $ZZ$) production.
\Tab{t:sampcomp} summarizes the expected sample compositions that are obtained by scaling to \invpb{955} from a previous \ttbar analysis with \invpb{318}~\cite{Abulencia:2005aj}. A detailed description of the background estimation is given in Ref.~\cite{Acosta:2004hw}.

\begin{table}
\caption[Event selection and sample composition]
{Event selection requirements for the \onetag and \twotag event samples.}
\label{t:evsel}
\begin{ruledtabular}
\begin{tabular}{ccc}
Event selection category & \onetag & \twotag \\
\hline
$\et^{e}$ (\gevnoarg) or $\pt^{\mu}$ (\gevcnoarg) & \multicolumn{2}{c}{$>20$} \\
\met (\gevnoarg) & \multicolumn{2}{c}{$>20$} \\
Jets 1--3 \et (\gevnoarg) & \multicolumn{2}{c}{$>15$} \\
Jet 4 \et (\gevnoarg) & $>15$ & $>8$ \\
Number of $b$ tags & $1$ & $\geq 2$ \\
\end{tabular}
\end{ruledtabular}
\end{table}

\begin{table}
\caption[Sample composition]
{The sources and expected numbers of background events, and the number of events observed for the \onetag and \twotag event samples in our \invpb{955} dataset. }
\label{t:sampcomp}
\begin{ruledtabular}
\begin{tabular}{ccc}
Background source & \onetag & \twotag\\
\hline
$W$ + jets & 21.4 $\pm$ 4.9 & 4.6 $\pm$ 1.5\\
non-$W$ & 7.0 $\pm$ 1.5 & 0.9 $\pm$ 0.2\\
EWK & 2.8 $\pm$ 0.6 & 0.17 $\pm$ 0.04\\
\hline
Total background events& 31.2 $\pm$ 7.0 & 5.7 $\pm$ 1.8 \\
\hline
Observed events & 171 & 82 \\
\end{tabular}
\end{ruledtabular}
\end{table}

After event selection, the analysis proceeds in three steps.  First, we reconstruct a top quark mass \mreco from each event using a kinematic fitter.  The width of the reconstructed mass distribution for the selected events is sensitive to \gtop.  The second step is a likelihood fit of the reconstructed mass distributions using simulated signal and background distributions that yields an estimator of \gtop.  Finally, we use a frequentist prescription (with Bayesian treatment of systematic uncertainties~\cite{Cousins:1991qz}) to determine a 95\%~C.L. upper limit on \gtop in the physically allowed region.

We perform a \chisq
minimization to fit the momenta of the \ttbar decay products and
determine \mreco for each event using the four leading jets. We assume that the final state arises from
the decay of a \ttbar pair into $W$ bosons and $b$ quarks. To resolve the ambiguity arising from the different ways of assigning the jets to the four quarks,
we require that $b$-tagged jets are assigned to $b$ quarks and select the assignment with the lowest \chisq. This kinematic fitter is used
in other CDF analyses and is described in detail in Ref.~\cite{Abulencia:2005aj}.
In the \chisq fit, both sets of $W$ decay daughters are constrained to have
the invariant mass of the $W$ boson, and both $Wb$ states are constrained to
have the same mass. Although the top and antitop quark will likely be produced with different masses, we confirmed that there is no significant difference in the sensitivity resulting from the $m_{t}=m_{\bar{t}}$ condition even for a large value of \gtop.

To distinguish between different values of \gtop, we compare the \mreco 
distribution from our data to a series of samples created using the {\sc pythia} 6.216 event generator~\cite{Sjostrand:2000wi} and a full detector simulation.  We use samples with $\mtop=\gevcc{175}$ and various values of \gtop between \gev{0.001} and \gev{100}.  
Although {\sc pythia} does not fully account for quantum interference with irreducible background diagrams and off-shell effects, for $\gtop~\lesssim~\gev{30}$ these effects are small, and the existing description is expected to be adequate~\cite{pythia_width_pc}.  \wjets background events are generated using {\sc alpgen} 1.3~\cite{r_alpgen}, with parton showering and fragmentation in {\sc herwig} 6.504~\cite{r_herwig}.
An unbinned extended maximum likelihood fit~\cite{Barlow:1990vc} is performed using parameterized signal and background \mreco templates. As an example, \fig{f:templates} shows templates for the signal \mreco distribution in the \twotag subsample at three values of \gtop. We parameterize the \mreco distributions as a function of \gtop. Small shifts in the mean of the templates are induced by the interplay between the top mass Breit-Wigner distribution and the parton distribution functions that preferentially produce events with low quark masses. In the likelihood fit, we constrain the background templates in the \twotag and \onetag samples to the levels given in \tab{t:sampcomp}. 
The best fit value \gtopfit is the width which maximizes the likelihood.
\begin{cfigure}
\includegraphics{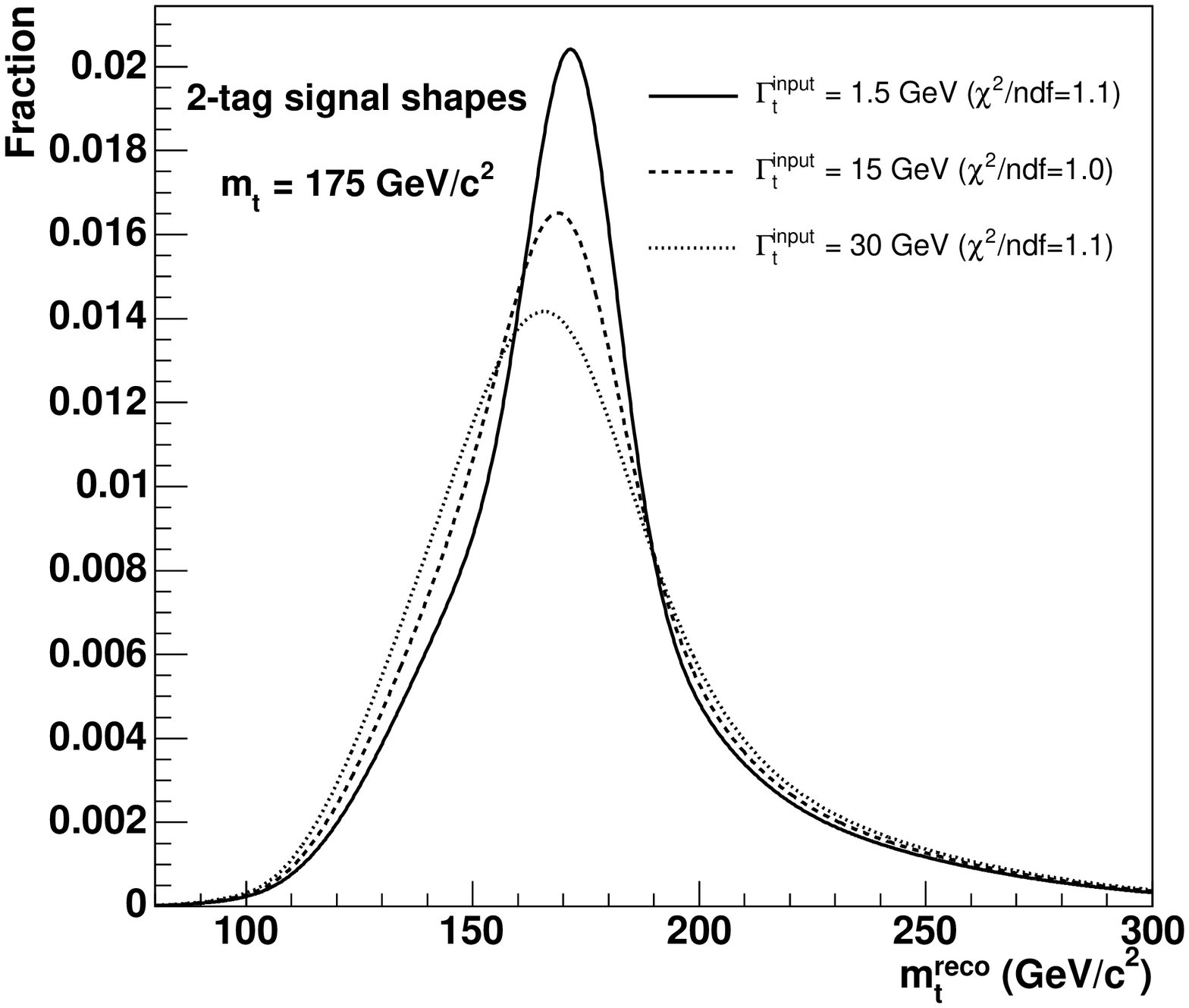}
\caption[Signal \mreco templates]
{The parameterized signal \mreco distributions and goodness of the parameterization are shown for the \twotag subsample at three different values of \gtop.  The parameterization is determined in a global fit to all the \twotag templates. The parameterized signal \mreco distributions for \onetag are similar.}
\label{f:templates}
\end{cfigure}

We allow negative \gtopfit values that represent \mreco distributions narrower than the nominal due to statistical fluctuations. The expected \mreco distribution for a negative \gtopfit is derived from an extrapolation of the parameterization to the unphysical region. The reconstructed top mass distributions from the data and the results of the likelihood fit are shown in \fig{f:datafits}. The data are consistent with the fitted curves with the preferred value of $\gtopfit=\gev{-4.8}$. For the data sample used in this analysis, we expect to measure a negative \gtopfit about 40\% of the time and \gtopfit less than \gev{-4.8} about 25\% of the time if the total width is \gev{1.5}. The limit on the true value of \gtop, however, will be constrained to the physical region.   

\begin{cfigure}
\includegraphics{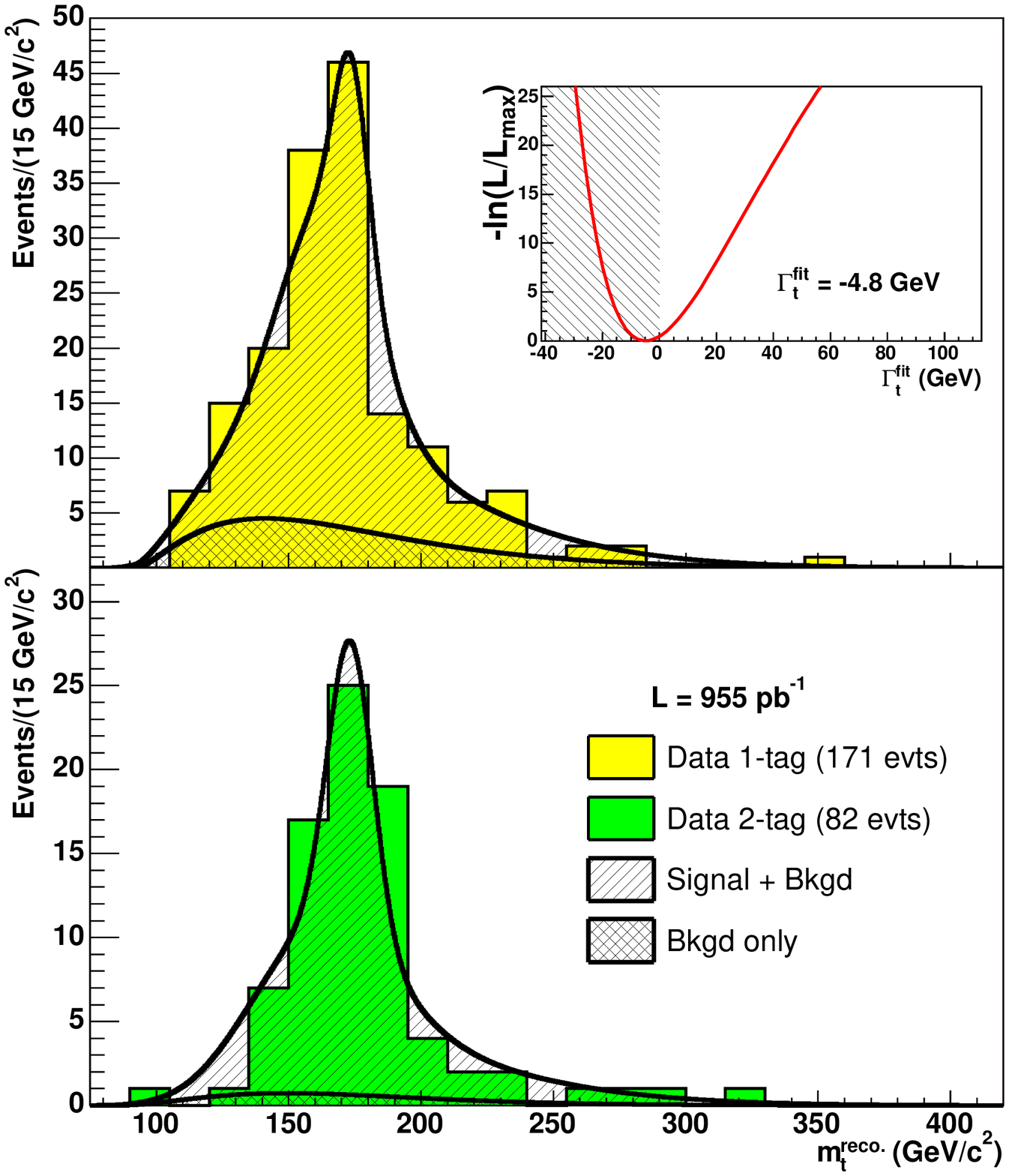}
\caption[Data distributions and likelihood fits]
{The \mreco distribution is shown for \onetag and \twotag data samples overlaid with the signal and background distributions from the combined fit.  A \gtopfit likelihood scan is shown in the inset; for the shaded non-physical region ($\gtopfit<0$), the likelihood uses \mreco distributions extrapolated from the parameterization to shapes with $\gtopfit>0$.}
\label{f:datafits}
\end{cfigure}

To set a limit on the true value of \gtop, we employ the Neyman construction~\cite{Neyman:1937} to ensure a coverage of at least 95\%.  The likelihood-ratio ordering principle due to Feldman and Cousins~\cite{Feldman:1997qc} provides a smooth transition from one-sided to two-sided limits and usually guarantees a non-empty interval.  We derive the confidence belts from ensembles of simulated experiments in which signal events are selected from the simulated samples generated with different values of \gtop.  The \gtopfit distribution from each such ensemble is convoluted with a shape that represents the effects of systematic uncertainties as described below.

Since the top quark mass reconstruction is dominated by the measurement of jet energy, and since our fit is largely determined by the peak and the width of the \mreco distribution, the uncertainties on the jet energy scale and the jet energy resolution are the dominant uncertainties in the top width measurement.  The uncertainties on the jet energy scale calibration are extensively studied using a combination of simulated and data control samples~\cite{r_NIMJES} and amount to about 3\% of the measured jet energy for jets in the \ttbar sample.  The effect on the \gtopfit distribution is non-linear, with larger \gtopfit being more likely for larger jet energy scale shifts in either direction.  This is because the only degree of freedom in the templates is the width, and a signal template with a larger \gtop accommodates the events with the shifted peak.

%We study the jet energy resolution by selecting events with one jet and one
%high-\pt photon and comparing their energies.  

%We determine the magnitude of a \sigunit{1} uncertainty on
%\gtopfit by adding 10\% -- 4\% Gaussian smearing per jet in signal Monte Carlo events.

We select events with one jet and one
high-\pt photon and compare their energies to study the jet energy resolution.
Data and {\sc pythia} events show similar jet resolution of 15\% -- 10\% for jet transverse energies between 
\gev{20} and \gev{200}, respectively. 
Taking into account statistical uncertainty of the data, we define a \pt-dependent systematic uncertainty on jet resolution of 10\% -- 4\% to cover the difference. Then, we add Gaussian smearing with corresponding uncertainty to each jet in signal Monte Carlo events.
We also study smaller systematic uncertainties in \gtopfit related
to the background \mreco shape, Monte Carlo statistics, the Monte Carlo
generator, the parton distribution functions, and other signal modeling
effects~\cite{Abulencia:2005aj}.  The combined convolution shape, accounting
for all systematic uncertainties, has a shift of \gev{-0.4} and an RMS of
\gev{4.6}.  This can be compared to the RMS from statistical effects only:
between \gev{6.6} and \gev{10.1} for simulated experiment ensembles using
\gtop of 1.5 -- \gev{30}. \Fig{f:fcbands} shows the 95\% confidence belts
after including systematic uncertainties.  The fitted value from data,
$\gtopfit=\gev{-4.8}$, corresponds to a limit of $\gtop<\gev{13.1}$ at 95\%~C.L.

Our measurement assumes a fixed value for the top quark mass of
\gevcc{175}. The one-dimensional likelihood is sensitive to this
assumption in the same way as described above for jet energy scale
uncertainties. In particular, if the top quark mass is consistent with the
current world average of \gevcc{\measErr{171.2}{2.1}}~\cite{PDBook},
the confidence belts would shift to higher \gtopfit values, resulting in an
upper limit of \gtop lower than what we quoted in this paper. 

In summary, using 253 top-antitop pair candidate events we present the first
direct experimental upper limit on the total decay width of the top quark,
$\gtop~<~\gev{13.1}$ at 95\%~C.L. for $\mtop~=~\gevcc{175}$.  This corresponds to a limit on the top
quark lifetime of $\tau_{t}~>~\genunit{5\times10^{-26}}{s}$.
This measurement is statistically limited and its dominant systematic uncertainties are likely reducible with statistics. The precision of this measurement, therefore, will continue to improve over the course of Run~II of the Tevatron.

\begin{cfigure}
\includegraphics{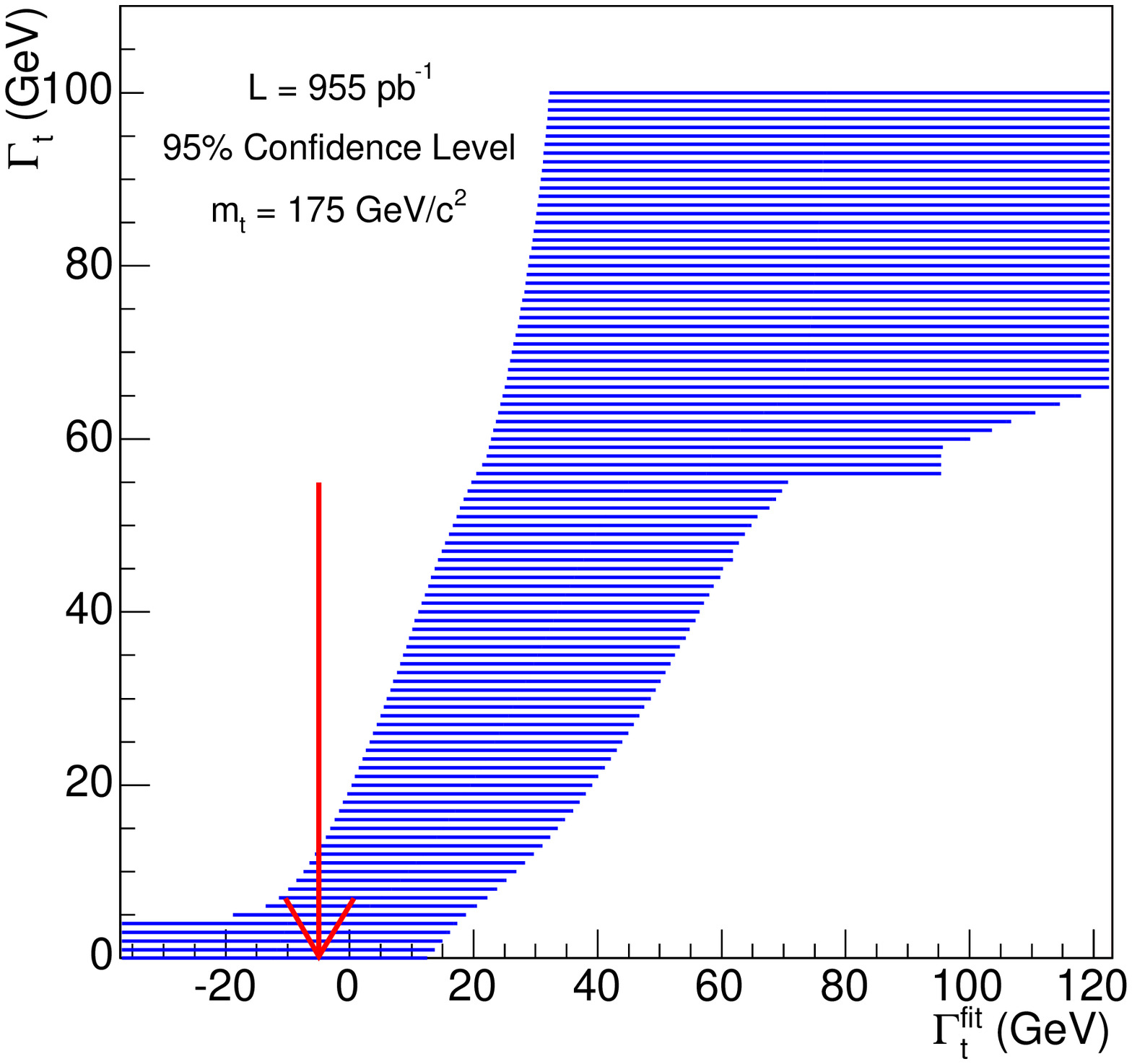}
\caption[Feldman-Cousins bands with data limit]
{The confidence band in \gtopfit for a 95\%~C.L. is shown.  Results from simulated experiments assuming a \invpb{955} dataset at different values of \gtop are convoluted with a smearing function to account for systematic uncertainties.  The fitted value from the data is indicated by an arrow.}
\label{f:fcbands}
\end{cfigure}

\begin{acknowledgments}
We thank Torbj\"orn Sj\"ostrand, Stephen Mrenna, Peter Skands, Alessandro Ballestrero, and Ezio Maina for discussions related to the Monte Carlo modeling of the top quark lineshape.
We thank the Fermilab staff and the technical staffs of the participating institutions for their vital contributions. This work was supported by the U.S. Department of Energy and National Science Foundation; the Italian Istituto Nazionale di Fisica Nucleare; the Ministry of Education, Culture, Sports, Science and Technology of Japan; the Natural Sciences and Engineering Research Council of Canada; the National Science Council of the Republic of China; the Swiss National Science Foundation; the A.P. Sloan Foundation; the Bundesministerium f\"ur Bildung und Forschung, Germany; the Korean Science and Engineering Foundation and the Korean Research Foundation; the Science and Technology Facilities Council and the Royal Society, UK; the Institut National de Physique Nucleaire et Physique des Particules/CNRS; the Russian Foundation for Basic Research; the Comisi\'on Interministerial de Ciencia y Tecnolog\'{\i}a, Spain; the European Community's Human Potential Programme; the Slovak R\&D Agency; and the Academy of Finland.
\end{acknowledgments}

\bibliographystyle{apsrev}
\bibliography{topwidthRefs}

\end{document}